\documentclass[reprint,amsmath,amssymb,preprintnumbers,aps,prd,nofootinbib]{revtex4-2}
\usepackage{graphicx}
\usepackage{comment}
\usepackage{booktabs}
\usepackage{array}
\usepackage{hyperref}
\usepackage{xcolor}
\usepackage{float}
\usepackage{siunitx}
\usepackage{bbm}
\usepackage{bbold}
\newenvironment{newtext}{\color{blue}}{}

\newenvironment{redtext}{\color{red}}{}
\newcommand{\red}[1]{\textcolor{red}{#1}}
\newcommand{\cmt}[1]{\textcolor{magenta}{\textbf{[#1]}}}

\begin{document}
\preprint{CERN-TH-2026-313}
\title{KEBBABS: Kilometric Enhanced Balloon-Baseline for Axion Birefringence Search}

\author{Tommaso Feri}
\email{tommaso.feri@phd.units.it}
\affiliation{Dipartimento di Fisica, Universit\`a di Trieste, Strada Costiera 11, 34151 Trieste, Italy}
\affiliation{INFN, Sezione di Trieste, Via A. Valerio 2, 34127 Trieste, Italy}

\author{Silvia Gasparotto}
\email{silvia.gasparotto@cern.ch}
\affiliation{CERN, Theoretical Physics Department, Esplanade des Particules 1, Geneva 1211, Switzerland}

%\author{Diego Blas}
%\email{dblas@ifae.es}
%\affiliation{Institut de Física d’Altes Energies (IFAE), The Barcelona Institute of Science and Technology, Campus UAB, 08193 Bellaterra (Barcelona), Spain}
%\affiliation{Institució Catalana de Recerca i Estudis Avançats (ICREA), Passeig Lluís Companys 23, 08010 Barcelona, Spain}

\date{\today}

\begin{abstract}
We propose a broadband search for ultralight axion dark matter through
the rotation of laser polarization induced by the axion--photon
coupling. Two stratospheric balloons flying at altitudes of 30--40 km
form a free-space optical link. One balloon sends a linearly polarized beam to the other, which reflects it back to a polarimeter on the first balloon. The long physical
baseline provides an alternative to the repeated reflections used in
resonant optical cavities. We target axion masses in the range
$4\times10^{-15}\,\mathrm{eV}\lesssim m_a\lesssim10^{-11}\,\mathrm{eV}$.
The lower edge corresponds to a frequency of about $1\,\mathrm{Hz}$,
below which gondola roll about the line of sight can mimic a
polarization signal.
At $m_a=10^{-14}\,\mathrm{eV}$, a low-cost Pathfinder using
flight-demonstrated technology could probe couplings as small as
$g_{a\gamma}\simeq2.8\times10^{-11}\,\mathrm{GeV}^{-1}$, about a
factor of two below the CAST bound. At the same mass, the Feasible
today configuration could probe couplings just below the astrophysical
bounds considered here, while the Futuristic configuration could reach
$g_{a\gamma}\simeq6.4\times10^{-15}\,\mathrm{GeV}^{-1}$, roughly two
orders of magnitude below those bounds.
\end{abstract}

\maketitle

% ============================================================
\section{Introduction}
% ============================================================

Axions and axion like particles are well-motivated extensions of the Standard Model and possible dark matter candidates~\cite{PecceiQuinn1977,Weinberg1978,Wilczek1978,Arvanitaki2010,Marsh2016,Sikivie1983,GrahamRajendran2013}. If axions couple to electromagnetism through the interaction $g_{a\gamma}a F_{\mu\nu}\tilde{F}^{\mu\nu}$, a time dependent axion background induces different phase velocities for the two circular polarizations of light, a phenomenon known as \textit{axion birefringence}~\cite{Carroll:1989vb,Carroll:1991zs,Harari:1992ea}. As a result, a linearly polarized beam acquires a small rotation of its polarization axis, which depends on the different field values at the emitter and receiver. For an oscillating axion dark matter background, this rotation oscillates at the axion Compton frequency, with an amplitude proportional to the local axion field. This alternating-current (AC) effect therefore provides a direct probe of the local axion dark matter field. See Sec.~III-(11) of Ref.~\cite{Arza:2026rsl} for a review of the sensitivity of this observable across different frequency ranges.

Existing laboratory proposals obtain a long interaction time primarily
through resonant optical cavities. DANCE, LIDA, and ADBC use dedicated
bow tie resonators~\cite{Obata2018,Oshima2023,LIDA2024,ADBC2019,ADBC2024},
while related searches have also been proposed using gravitational wave
detectors~\cite{Nagano2019,Nagano2021,KAGRA2021,MartynovMiao2020,Michimura2024, Gue:2024txz}.
Resonant enhancement enables a compact instrument, but the same repeated round trips that build up the axion signal can also amplify polarization systematics and technical noise, requiring stringent control of cavity and laser stability, as demonstrated by the first runs of DANCE Act-1~\cite{Oshima2023}.

A natural alternative is to exploit much longer baselines, as proposed, for example, by adapting the inter spacecraft links of LISA~\cite{LISAaxion2025,LISAmission}. While such baselines offer excellent potential sensitivity, this approach would require substantial modifications to an already planned space mission~\cite{Gue:2024txz}, with correspondingly high costs associated with satellite based instrumentation.

Here we consider an intermediate platform, a folded free space laser link between two stratospheric balloons. Scientific balloons can operate above 99$\%$ of the atmosphere, corresponding of an altitude of 40 km, where the residual
air column along the link has a negligible impact on the polarization
measurement (see Appendix~\ref{app:atm}). They are a well established platform for atmospheric science and Earth observation, as well as for fundamental physics and astrophysics, with examples including measurements of cosmic microwave background temperature anisotropies with BOOMERANG~\cite{Boomerang:2000efg} and polarization with SPIDER~\cite{SPIDER:2021ncy}, searches for cosmic antimatter with BESS-Polar~\cite{BESS-Polar}, observations of Galactic positron annihilation radiation with COSI~\cite{COSI}, and galaxy cluster imaging for weak gravitational lensing studies with SuperBIT~\cite{SuperBIT2024}.
A similar two balloon link has recently been proposed for free space quantum communication~\cite{Raja_baloon}, relying on comparable technology.

%The proposal builds on subsystems with relevant experimental heritage, precision polarimetry~\cite{PVLAS,Zavattini2018,Kemp1969}, balloon borne pointing systems~\cite{SuperBIT2024,SunriseIII}, and optical links to and between balloons~\cite{STROPEX,BalloonFT}, including bidirectional laser communication between balloons over distances of up to 107 km achieved by Project Loon~\cite{Loon}. Developing such links could also benefit free space quantum communication between balloons, which relies on similar terminals and pointing systems and, for polarization encoding, on the same characterization of polarization errors that our measurement performs~\cite{Raja_baloon}.

In this paper, we consider three configurations corresponding to increasing levels of technological maturity: a Pathfinder, built from already demonstrated subsystems, a Feasible today configuration, combining the best currently demonstrated subsystems, and a Futuristic configuration, based on projected technology.
In the Pathfinder configuration, Balloon A hosts the laser and polarimetric
readout, while Balloon B carries a mirror turnaround that redirects the beam
back to Balloon A. The polarization is therefore prepared and measured on the
same platform, ensuring a common local reference frame for both polarization
generation and detection. In the two more advanced configurations, both
balloons host an active laser polarimeter payload enabling bidirectional
measurements.  This produces two reciprocal measurements,
$A\!\to\!B\!\to\!A$ and $B\!\to\!A\!\to\!B$, over the same physical baseline.
Their coherent average is the science channel, while their difference is an
internal null channel for diagnosing local instrumental disturbances. 

%The interaction time is obtained from physical distance rather than repeated circulation in a resonator.  In the long wavelength regime the axion response is proportional to the one way separation $L$, so a long baseline is intrinsically favorable. with relevant experimental heritage, including precision polarimetry~\cite{PVLAS,Zavattini2018,Kemp1969}, stratospheric opticallinks~\cite{STROPEX,Loon,BalloonFT}, and pointing systems~\cite{SuperBIT2024,SunriseIII}.

%At the reference mass $m_a=10^{-14}\,\mathrm{eV}$ we find that the Pathfinder reaches $g_{a\gamma}\simeq3.0\times10^{-11}\,\mathrm{GeV}^{-1}$, about a factor of two below the CAST bound, the Feasible today configuration $6.6\times10^{-13}\,\mathrm{GeV}^{-1}$, comparable to astrophysical bounds, and the Futuristic configuration $6.4\times10^{-15}\,\mathrm{GeV}^{-1}$, well below the astrophysical bounds over the whole low mass part of its response.
%Long polarization-preserving free space links are also directly relevant to stratospheric quantum communication~\cite{Raja_baloon,QKDturb}, providing an independent technology motivation.

In Sec.~\ref{sec:signal}, we derive the axion-induced polarization rotation along the
round-trip optical path and show how the balloon separation sets the
response as a function of axion mass. Section~\ref{sec:concept} describes the optical
design and the main contributions to the noise budget. In Sec.~\ref{sec:reach}, we
present the projected sensitivity of the three configurations to the
axion--photon coupling and compare it with existing laboratory and
astrophysical bounds. Section~\ref{sec:summary} summarizes our findings and outlines
the next steps toward a Pathfinder flight.

% ============================================================
\section{Axion signal and baseline response}\label{sec:signal}
% ============================================================

The axion--photon interaction,
\begin{equation}
\mathcal{L}\supset-\frac{g_{a\gamma}}{4}aF_{\mu\nu}\widetilde F^{\mu\nu},
\end{equation}
couples the electromagnetic field to the pseudoscalar axion field. Since
$F_{\mu\nu}\widetilde F^{\mu\nu}$ is odd under parity transformations ($\Vec{x} \xrightarrow{} -\Vec{x}$, where $F \xrightarrow{} F$ and $\tilde{F} \xrightarrow{} -\tilde{F}$), while the axion transforms as
$a\to-a$, the full interaction is parity invariant. However, a nontrivial axion
background provides a parity-odd environment for electromagnetic propagation. In
particular, a time dependent axion field modifies Maxwell's equations in such a
way that the two circular polarizations of light propagate differently.

In the geometric-optics regime, for an axion background varying slowly compared
with the photon wavelength, the two circular polarizations propagate with phase
velocities
\begin{equation}\label{eq:phasevelocity}
c_{L/R}\simeq1\mp\frac{g_{a\gamma}\dot a}{2k},
\end{equation}
where $k$ is the photon wavenumber~\cite{Carroll:1989vb,Harari:1992ea}.
The resulting difference in the accumulated phases of the two helicities produces
a rotation of the polarization axis of a linearly polarized wave. For a ray emitted
at time $t_0$ and detected at time $t_1$, the rotation angle is
\begin{equation}
\theta=\frac{g_{a\gamma}}{2}\left[a(t_1)-a(t_0)\right].
\label{eq:endpoint}
\end{equation}
Remarkably, the resulting signal depends only on the values of the axion field at
the endpoints of the light trajectory, rather than on its detailed profile along
the propagation path. A derivation is provided in Appendix~\ref{app:signal}.

For axion dark matter we write
\begin{equation}
a(t)=a_0\cos(\omega_a t+\delta),\qquad
a_0=\frac{\sqrt{2\rho_{\rm DM}}}{m_a},
\label{eq:afield}
\end{equation}
with $\omega_a=m_ac^2/\hbar$. Let $L$ denote the one-way balloon separation
and $x\equiv\omega_aL/c$. A single reflection reverses the photon helicity and
causes the outbound and return contributions to partially cancel. The remote
terminal therefore uses an even number of reflections. For this even-reflection geometry,
\begin{equation}
\theta(t)=-g_{a\gamma}a_0\sin x\,\sin(\omega_at+\delta+x),
\end{equation}
so the rotation amplitude is
\begin{equation}
\theta_0(m_a)=g_{a\gamma}\frac{\sqrt{2\rho_{\rm DM}}}{m_a}F_L(m_a),
\label{eq:response}
\end{equation}
with the baseline response function
\begin{equation}
F_L(m_a)=\left|\sin\!\left(\frac{\omega_aL}{c}\right)\right|.
\label{eq:FL}
\end{equation}
The response therefore forms a comb of maxima and nulls. Its first maximum
occurs when the round-trip travel time is approximately half an axion period,
\begin{equation}
\nu_{\rm opt}\simeq\frac{c}{4L}=\frac{c}{2L_{\rm path}},
\qquad L_{\rm path}=2L,
\end{equation}
which is $2.7$ kHz, $330$ Hz and $112$ Hz for the three configurations
considered below.
Changing the balloon separation shifts the peaks and zeros of the frequency response while leaving the polarimetric readout unchanged. As the balloons drift, their separation $L(t)$ varies slowly, as illustrated by the Project
Loon GPS data in App.~\ref{app:separation}. We account for this variation
by defining the root-mean-square (RMS) response over the observation,
\begin{equation}
    F_L^{\rm rms}(m_a)
    = \sqrt{\left\langle |F_{L(t)}(m_a)|^2
      \right\rangle_{\rm obs}},
\end{equation}
where the brackets denote a time average over the periods when the optical
link is operating. A frequency at which the response vanishes for one
separation generally has a finite response at other separations. The drift
therefore turns the fixed-baseline zeros at nonzero frequencies into
finite minima.

% ============================================================
\section{Experimental concept and noise budget}\label{sec:concept}
% ============================================================

% REV3: this section had the label "sec", which collided with the reference in
% REV3: Appendix B 3 a and produced an undefined cross-reference.
\subsection{Optical architecture}\label{sec:arch}

\begin{figure*}[t]
\centering
\includegraphics[width=0.95\textwidth]{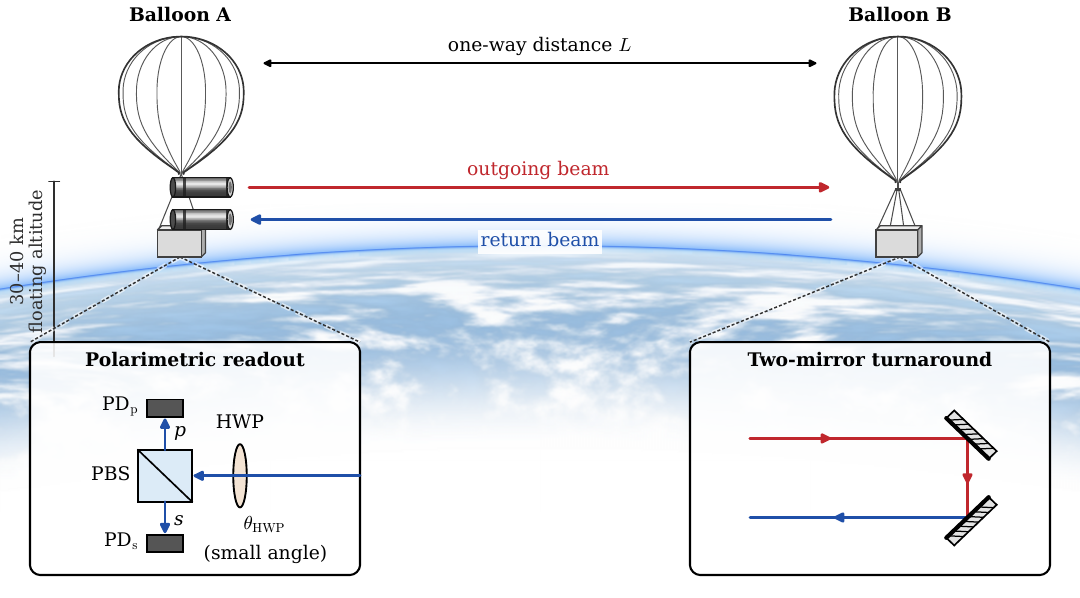}

\caption{One-way optical architecture.  The drawing shows one round-trip channel:
a laser source placed on balloon A send the polarized beam towards ballon B where its reflected by the two mirror turnaround and redirected to the polarimeter on ballon A.  The terminal on ballon A carries a 
transmit telescope, which sets the divergence of the outgoing beam, and a
receive telescope, which collects the return. In the
Feasible today and Futuristic benchmarks the same hardware is installed on
both balloons, producing the reciprocal channel in the opposite direction.
The two measurements form the science and null combinations of
Eq.~\eqref{eq:sciencenull}. Each returning beam is analyzed with a half-wave
plate, a polarizing analyzer, and two photodetectors measuring the $s$ and $p$
outputs, as detailed in Appendix~\ref{app:polarimeter}.}
\label{fig:setup}
\end{figure*}

The optical path is shown in Fig.~\ref{fig:setup}. The active terminal
on balloon A carries two separate telescopes, one for transmission and one for
reception. A linear polarizer prepares the outgoing beam before it enters the transmit telescope. 
The transmit telescope, with diameter $D_{\rm T}$, expands the
outgoing beam and reduces its diffraction-induced divergence. For a uniformly
illuminated circular aperture, the angular spread of the transmitted beam, measured from the optical axis, is approximately
$\theta_{\rm diff}=1.22\lambda/D_{\rm T}$, where $\lambda$ is the laser
wavelength and the factor $1.22$ arises from the diffraction pattern of a
circular aperture. Thus, at fixed wavelength and pointing performance,
increasing the transmit aperture reduces the spreading of the beam over the
propagation distance.

At balloon B, the beam is reflected by two flat mirrors at $90^\circ$ to each
other, each ideally forming an angle of $45^\circ$ with the incoming laser
beam. This arrangement, known as a roof mirror, is the simplest setup
providing the even number of reflections required to preserve the axion
induced polarization signal (App.~\ref{app:parity}).

We consider the transmitted laser to be continuously pointed from balloon A
towards the remote terminal on balloon B~\footnote{In Loon’s optical flight tests, each balloon used GPS and onboard sensors to aim its telescope toward the other balloon. A camera then tracked the other balloon’s guide laser, and a motorized mount kept the spot centered as the gondolas moved~\cite{Loon}. In the bidirectional KEBBABS configurations, each balloon could use the incoming science beam from the other balloon as its guide light, without a separate laser control.}

We denote by $\sigma_p$ the root
mean square (RMS) residual angular fluctuation after pointing control, which
quantifies the typical pointing error that remains uncorrected. Similarly, we
assume that the orientation of the roof mirror on balloon B is actively
controlled so that the beam is redirected towards balloon A, with RMS residual
error $\varepsilon_{\rm turn}$.\footnote{ As shown in the Appendix, the two angular errors contribute with
the same geometric weight to the beam broadening in Eq.~\eqref{eq:wret}.
A broader beam delivers less optical power to the receiver, increasing
the relative photon shot noise.}

After reflection, the returning beam is collected by a second telescope of
diameter $D_{\rm R}$ and directed towards the polarimetric readout described in
Sec.~\ref{sec:readout}. The three apertures therefore perform distinct
functions: $D_{\rm T}$ is the telescope aperture that controls the diffraction induced divergence of the
outgoing beam, $D_{\rm turn}$ is the aperture of the mirrors that intercept the outgoing
beam at the remote balloon, and $D_{\rm R}$ is the telescope aperture that determines the fraction of the
returning beam collected at the active terminal.

The setup described here corresponds to a single round trip optical channel. In
the Feasible today and Futuristic configurations, the same architecture is
duplicated in the opposite direction. Denoting by
$\widehat\theta_{A\to B\to A}$ and $\widehat\theta_{B\to A\to B}$ the two
reciprocal estimates, we define

\begin{equation}
\widehat\theta_\pm=\frac{1}{2}
\left(\widehat\theta_{A\to B\to A}\pm\widehat\theta_{B\to A\to B}\right),
\label{eq:sciencenull}
\end{equation}
where $\widehat\theta_+$ is the science channel and $\widehat\theta_-$ is a
null channel that contains no axion signal and is used to identify and control
instrumental systematic errors. \begin{comment}
If the noise of the two lines is uncorrelated,
the science channel has
\begin{equation}
 S_{\theta,+}=\frac{S_{\theta,{\rm line}}}{2},
 \label{eq:crossspectra}
\end{equation}
i.e.\ a $\sqrt2$ reduction in amplitude spectral density with respect to a
single line, while a disturbance common to the two lines enters
$\widehat\theta_+$ with full weight and is absent from $\widehat\theta_-$. Throughout our optical link model, we approximate the transmitted
and returning beams as Gaussian beams; the validity and limitations
of this approximation are discussed in
Appendix~\ref{app:geo}. 
\end{comment}

\subsection{Polarimetric readout}\label{sec:readout}

To measure the polarization rotation of the returning beam, we adopt the
readout implemented in DANCE Act-1~\cite{Oshima2023}. Its derivation is given
in Appendix~\ref{app:polarimetric-readout}; here we summarize only the
expressions needed for the signal reconstruction.

The linear polarizer placed before the transmit telescope prepares the
outgoing beam in a linear polarization aligned with the polarizing analyzer at
the end of the readout chain. We denote by $s$ this polarization,
perpendicular to the plane of incidence of the analyzer, and by $p$ the
orthogonal one, lying in that plane. Since the polarizer and the analyzer are
on the same platform, they share a common reference frame: in the absence of a
signal the returning beam is entirely in $s$, and a polarization rotation
$\theta(t)$ appears as a small $p$ component.

Measured directly, this component would give a power
$P_p\simeq P_{\rm rx}\,\theta^2$, quadratic in an already tiny rotation. To
obtain a linear response, a half wave plate placed before the analyzer
rotates the polarization by a small and known angle
$\epsilon=2\theta_{\rm HWP}$, where $\theta_{\rm HWP}$ is the angle between its
fast axis and $s$. This transfers a small reference field into $p$, which
interferes with the component generated by the signal. The two analyzer
outputs are
\begin{equation}
P_p(t)\simeq P_{\rm rx}(t)\left[\epsilon+\theta(t)\right]^2,
\qquad
P_s(t)+P_p(t)\simeq P_{\rm rx}(t),
\label{eq:readout-powers}
\end{equation}
where $P_{\rm rx}(t)$ is the total optical power reaching the polarimeter after
the round trip, computed in Appendix~\ref{app:link}. For $|\theta|\ll\epsilon$,
\begin{equation}
P_p(t)\simeq P_{\rm rx}(t)\left[\epsilon^2+2\epsilon\,\theta(t)\right],
\end{equation}
so the signal produces a power variation linear in the rotation,
\begin{equation}
\Delta P_p(t)\simeq 2\epsilon\,P_{\rm rx}(t)\,\theta(t),
\end{equation}
larger than the quadratic term $P_{\rm rx}\theta^2$ by a factor
$2\epsilon/\theta\gg1$.

Since $\epsilon$ is independently calibrated, the rotation is reconstructed
from the two measured powers as
\begin{equation}
\widehat\theta(t)=
\sqrt{\frac{P_p(t)}{P_s(t)+P_p(t)}}-\epsilon
\simeq
\frac{\Delta P_p(t)}{2\epsilon\,P_{\rm rx}(t)}.
\label{eq:estimator-main}
\end{equation}
The normalization to the total received power removes fluctuations of the
laser power and of the link transmission that are common to the two ports.
The readout therefore requires only standard polarization optics and two
photodetectors.

\subsection{Noise model}\label{sec:noise}

We model the noise affecting the polarimetric measurement with two contributions,
\begin{equation}
 S_{\theta}(f) = S_{\theta,{\rm shot}}(f) + S_{\theta,{\rm tech}}(f).
 \label{eq:line-noise}
\end{equation}

The first term is photon shot noise, i.e. the Poisson fluctuation of the
number of detected photons. It depends only on the optical power
reaching the detector, and can be written as
\begin{equation}
 S_{\theta,{\rm shot}}^{1/2}
 = \sqrt{\frac{hc}{2\,\mathrm{QE}\,P_{\rm rx}\,\lambda}},
 \label{eq:shot-main}
\end{equation}
where $h$ is the Planck constant, $c$ the speed of light, $\lambda$ the laser
wavelength, $P_{\rm rx}$ the optical power collected at the receiver, and
$\mathrm{QE}$ the quantum efficiency of the photodetector, i.e.\ the fraction
of incident photons converted into photoelectrons; we take
$\mathrm{QE}=0.8$ at $1550$ nm.
The returned power $P_{\rm rx}$ is set by the laser power, the three optical
apertures, the pointing accuracy of the two terminals, and the losses of the
optical components and of the free space link. The full calculation is presented in
App.~\ref{app:geo}.

The second term is the technical noise. The values adopted for each
technological level are motivated by the technical noise of existing
polarimetric measurements, as discussed in App.~\ref{app:tech}.

The residual interaction of the beam with the atmosphere could in principle
produce additional spurious rotations or a degradation of the signal, through
Faraday rotation and scattering, respectively. 
%As shown inAppendices~\ref{app:faraday} and~\ref{app:depol}, both effects are negligible in our case and are not included in the noise budget. 
In the geometry considered, Faraday rotation cancels because
the retraced path contributes a rotation equal in magnitude and opposite in
sign to that of the outgoing path (see App.~\ref{app:faraday}). Atmospheric
scintillation modulates the received optical power, but it does so equally in
the $p$ and $s$ ports; it therefore cancels at first order in the normalized
two-port rotation estimator~\eqref{eq:estimator-main}. Moreover ,in the bidirectional
configurations, the null channel provides an additional check against
disturbances that affect the two round-trip measurements differently.
Finally, roll of the gondolas about the line of sight is expected to dominate
below $\sim 1\,$Hz (App.~\ref{app:tech-roll}); we therefore adopt
$1\,$Hz as the lower frequency cut-off of our analysis. 

The two terms are uncorrelated and add in quadrature, so the total amplitude
spectral density (ASD) is 
\begin{equation}\label{eq:ASD}
    S_\theta^{1/2}=(S_{\theta,{\rm shot}}+S_{\theta,{\rm
tech}})^{1/2}.
\end{equation}

The values adopted for the three configurations are listed in
Table~\ref{tab:config}, together with the experimental components from which the
shot-noise term follows.
%%%%%%%
\subsection{Technological levels}\label{sec:levels}

Table~\ref{tab:config} defines three benchmark configurations corresponding to increasing levels of technological maturity. For each configuration, the laser power, pointing precision, fixed optical throughput, technical polarization noise, integration time are chosen to be consistent with the relevant demonstrated subsystem capabilities. Once these quantities are specified, the target inter-balloon distance $L$ and the
three aperture diameters are determined through the bounded optimization
described in Appendix~\ref{app:optimization}. Due to the balloon motion the actual distance between them fluctuates
around this target during the campaign, we treat these fluctuations in
Appendix~\ref{app:separation}. The optimization ranges are chosen to be technologically plausible for each configuration.
\begin{table*}[t]
\centering
\scriptsize
\begin{tabular}{@{}lcccc@{}}
\hline
Parameter & Pathfinder & Today & Futuristic & Ref. \\
\hline
Round-trip lines
& 1 & 2 & 2 & -- \\

Wavelength $\lambda$ (nm)
& 1550 & 1550 & 1550 & -- \\

Float altitude $h_{\rm float}$ (km)
& 30 & 35 & 40
& \cite{NASABalloonCapabilities} \\

Laser power $P_{\rm tx}$ (W)
& 0.10 & 2.0 & 100
& \cite{STROPEX,Roubal2025,Huang2022} \\

Pointing precision $\sigma_p$ ($\mu$rad)
& 8.0 & 1.8 & 0.5
& \cite{Loon,SuperBIT2024,Zhang2020} \\

Optical throughput $\eta_{\rm opt}$
& 0.86 & 0.86 & 0.92
& App.~\ref{app:etafixed} \\

Integration time $T$ (d)
& 30 & 50 & 365
& \cite{SuperBIT2024,NASASPB} \\

Target baseline $L$ (km)
& 27.5 & 226.4 & 667.0
& \cite{Loon} \\

Transmit aperture $D_{\rm T}$ (m)
& 0.39 & 1.13 & 1.50
& \cite{SuperBIT2024,SunriseIII,BLAST} \\

Receive aperture $D_{\rm R}$ (m)
& 0.50 & 1.30 & 1.50
& \cite{SuperBIT2024,SunriseIII,BLAST} \\

Mirror aperture $D_{\rm turn}$ (m)
& 0.50 & 1.30 & 1.50
& \cite{SuperBIT2024,SunriseIII,BLAST} \\

Returned power $P_{\rm rx}$ (W)
& $2.74\times10^{-3}$ & $1.41\times10^{-1}$ & $10.5$
& Eq.~\eqref{eq:prx-link} \\

Shot-noise ASD ($\mathrm{rad}/\sqrt{\mathrm{Hz}}$)
& $5.41\times10^{-9}$ & $7.55\times10^{-10}$ & $8.73\times10^{-11}$
& Eq.~\eqref{eq:shot-main} \\

Technical ASD ($\mathrm{rad}/\sqrt{\mathrm{Hz}}$)
& $5.0\times10^{-9}$ & $5.0\times10^{-10}$ & $5.0\times10^{-11}$
& App.~\ref{app:tech} \\
\hline
\end{tabular}

\caption{Benchmark configurations corresponding to the three technological
levels considered in this work. The references in the last column identify
the demonstrated capabilities used to motivate the parameter ranges. Once the
subsystem quantities are specified, the target one-way baseline and the three
aperture diameters are determined through the constrained optimization
described in Appendix~\ref{app:optimization}, which includes the drift of the
balloon separation. For the geometric parameters,
the cited experiments establish the plausibility of the relevant distance and
aperture scales rather than the prior demonstration of the complete optimized
configurations. In particular, the $667$ km baseline is a projection for the
Futuristic configuration.
The shot-noise and technical ASDs refer to a single round-trip line; the
returned power is evaluated at the optimized geometry.}
\label{tab:config}
\end{table*}
The \emph{Pathfinder} configuration is intended to demonstrate the complete measurement scheme at low cost while approaching sensitivity at the CAST scale. Its parameters are therefore based on capabilities already demonstrated in flight. We assume a $0.1$ W transmitter at $1550$ nm, a power level already demonstrated on stratospheric balloons~\cite{STROPEX}, while space qualified laser transmitters have reached powers of several watts~\cite{Roubal2025}. We adopt an RMS pointing error of $\sigma_p = 8\,\mu$rad. Pointing and tracking between two stratospheric balloons over separations exceeding $100$ km has already been demonstrated~\cite{Loon}. High pointing precision towards astronomical sources has also been achieved by balloon-borne telescopes, with SuperBIT demonstrating an RMS pointing stability of $1.65\,\mu$rad~\cite{SuperBIT2024}. Pointing accuracies of $0.5$--$1.0\,\mu$rad have also been demonstrated in satellite to ground optical links by the Micius satellite~\cite{Zhang2020}. Although these experiments involve different pointing geometries and operating conditions, their demonstrated performance, together with Loon's successful inter-balloon tracking~\cite{Loon}, supports the technological plausibility of our pointing requirement.
Given these assumptions, the optimization selects a $27.5$ km baseline, a $0.40$ m transmit aperture, and $0.50$ m receive and turnaround apertures, the latter two at the upper bound of this level. These dimensions remain at or below the $0.5$ m aperture of Super BIT~\cite{SuperBIT2024}.
The Pathfinder baseline is short enough that the curvature of the Earth lowers
the line of sight at mid-path by only $\simeq15$ m below the
float altitude (App.~\ref{app:atm}). A float altitude of $30$ km is therefore sufficient to keep
atmospheric effects negligible along the whole path.

The \emph{Feasible today} configuration combines more advanced subsystem capabilities without the Pathfinder low-cost constraint. We assume a $2$ W laser~\cite{Roubal2025}, and a pointing requirement of $1.8\,\mu$rad~\cite{SuperBIT2024}. The $50$ days of cumulative integration are compatible with the durations achieved by super pressure balloon campaigns~\cite{NASASPB}. 
The optimized geometry consists of $226$ km baseline, a $1.13$ m transmit aperture, and $1.30$ m receive and turnaround apertures. Over such a baseline,
a float altitude of $35$ km, close to that of NASA super pressure
balloons~\cite{NASABalloonCapabilities}, is sufficient, the line of sight at mid-point is 
$1$ km below float altitude, and atmospheric extinction over the $450$ km round trip
remains negligible (App.~\ref{app:atm}).

The \emph{Futuristic} configuration assumes more demanding performance, combining technologies that have not yet been all demonstrated on a stratospheric platform. We adopt a $100$ W laser, a power level already achieved in laboratory systems near $1550$ nm with linearly polarized, single-frequency output~\cite{Huang2022}. The pointing requirement of $0.5\,\mu$rad between balloons remains a technological challenge, however it's comparable to the pointing accuracies achieved by Micius in orbit~\cite{Zhang2020}. The assumed one year integration time could be also achieved over multiple campaigns rather than a single uninterrupted flight.
The optimized geometry has a $667$ km baseline and $1.5$ m transmit, receive,
and turnaround apertures, kept below the $2$ m aperture flown by
BLAST~\cite{BLAST}. We assume a float altitude of $40$ km, at this altitude the beam
stays above $30$ km the most of the time (App.~\ref{app:atm}).

\begin{figure*}[t!]
\centering
\includegraphics[width=0.85\textwidth]{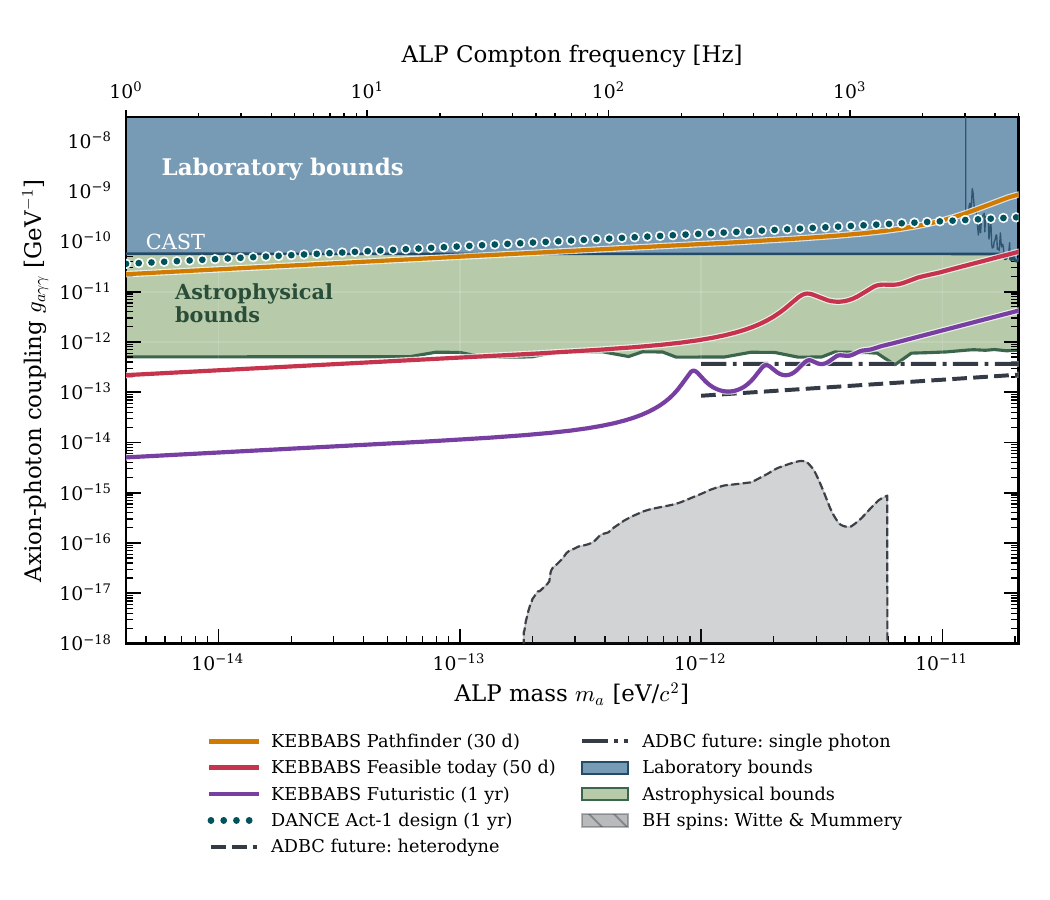}
\caption{Projected KEBBABS sensitivity to the ALP–photon coupling \(g_{a\gamma\gamma}\) as a function of the ALP mass. The upper axis gives the corresponding Compton frequency limited to 1 Hz, as lower values could be affected by the gondola motion \ref{app:tech-roll}. Solid curves show the nominal \(\mathrm{SNR}=1\) reach which include the gondola drift through the averaged baseline response, see App.~\ref{app:reach}. Blue shading indicates existing laboratory bounds, while green shading follows the lower envelope of astrophysical constraints~\cite{AxionLimits}. The dotted curve shows the final design of DANCE Act-1~\cite{Oshima2023}; dashed and dash–dotted curves show the ADBC heterodyne and single-photon projections\cite{ADBC2024}, respectively. The gray region shows the black hole superradiance bound on $f_a$~\cite{Witte:2024drg}, translated into $g_{a\gamma}$ via $|g_{a\gamma}| = \alpha_{\rm em}|C_{a\gamma}|/(2\pi f_a)$ with $C_{a\gamma}=1$.  }

\label{fig:reach}
\end{figure*}
% ============================================================
\section{Projected sensitivity}\label{sec:reach}
% ============================================================

We translate the polarization-rotation sensitivity of the three
configurations into a projected sensitivity to the axion--photon
coupling. Let $\sigma_\theta=\sqrt{S_\theta(\nu_a)}$ denote the
one-sided noise amplitude spectral density of the science channel at
the axion frequency $\nu_a$. In the Feasible today and Futuristic configurations, averaging the two reciprocal measurements reduces $\sigma_\theta$ by a factor of $\sqrt{2}$ for equal, uncorrelated noise Eq.~\eqref{eq:sciencenull}.

The axion field has a finite coherence time $\tau_c\simeq Q/\nu_a$,
with $Q\sim10^6$~\cite{Centers2021,Budker2014}. As derived in
App.~\ref{app:reach}, the RMS rotation amplitude corresponding to a signal-to-noise of unity (SNR) for an observation of duration $T$ is
\begin{equation}
\delta\theta_0 =
\begin{cases}
\sigma_\theta\sqrt{2/T}, & T\leq\tau_c,\\[3pt]
\sigma_\theta\sqrt{2/\tau_c}\,(\tau_c/T)^{1/4}, & T>\tau_c.
\end{cases}
\label{eq:ampunc}
\end{equation}
For $T\leq\tau_c$, the field remains coherent throughout the
observation, and the amplitude sensitivity scales as $T^{-1/2}$.
For $T>\tau_c$, we model the observation as $N=T/\tau_c$
statistically independent segments of duration $\tau_c$ and combine
their measured powers. The uncertainty on the mean power then scales
as $N^{-1/2}$. Since the signal power is proportional to the square
of the rotation amplitude, the amplitude sensitivity scales as
$N^{-1/4}$.

Combining the optical response in Eq.~\eqref{eq:response} with the
amplitude sensitivity in Eq.~\eqref{eq:ampunc} gives the nominal reach in the axion--photon coupling, corresponding to $\mathrm{SNR}=1$
\begin{equation}
g_{a\gamma}(m_a)=
\frac{m_a\,\delta\theta_0}
     {\sqrt{2\rho_{\rm DM}}\,F^{\rm rms}_L(m_a)},
\label{eq:reach}
\end{equation}
where we adopt a local dark matter density of
$\rho_{\rm DM}=0.4\,\mathrm{GeV\,cm^{-3}}$~\cite{Read2014}. 

Figure~\ref{fig:reach} compares the three KEBBABS configurations with
existing bounds and projected optical cavity sensitivities, starting at
$\nu_a=1$ Hz ($m_a\simeq4.1\times10^{-15}$ eV), below which the pendulation
of the flight train could mimic the axion induced polarization rotation
(App.~\ref{app:tech-roll}). 

At
$m_a=10^{-14}\,\mathrm{eV}$, the Pathfinder would probe
$g_{a\gamma}\simeq2.8\times10^{-11}\,\mathrm{GeV}^{-1}$, about a factor
of two below the CAST helioscope bound. The Feasible today configuration would reach
$g_{a\gamma}\simeq2.7\times10^{-13}\,\mathrm{GeV}^{-1}$, 
below the existing astrophysical bounds
at this mass and until $m_a\simeq10^{-13}\,\mathrm{eV}$~\cite{AxionLimits}. The Futuristic
configuration would probe couplings as small as
$6.3\times10^{-15}\,\mathrm{GeV}^{-1}$, extending a direct optical
search for axion dark matter to couplings roughly 80 times weaker than
those currently constrained by astrophysical observations. The Futuristic benchmark comes closest to the region constrained through axion superradiance~\cite{Witte:2024drg}.
We translate the resulting bound on the axion self-interaction scale $f_a$
into a bound on the photon coupling using
$|g_{a\gamma}|=\alpha_{\rm em}|C_{a\gamma}|/(2\pi f_a)$ with $C_{a\gamma}=1$.

% ============================================================
\section{Summary and outlook}\label{sec:summary}
% ============================================================

We have introduced KEBBABS, a new experimental approach to the search for ultralight axion dark matter through the polarization rotation of laser light exchanged between two stratospheric balloons. The proposed architecture offers an alternative to resonant optical
cavities: a long, tunable free space baseline provides the optical path,
while the two mirror turnaround limits the number of reflections at which
mirror birefringence can affect the polarization. The bidirectional configuration adopted in the Feasible Today and Futuristic benchmarks further enhances sensitivity and provides an internal null test.

Our sensitivity forecasts, based on three benchmarks spanning different levels of technological maturity, illustrate the potential of stratospheric optical links as a platform for axion dark matter searches. The optimized configurations are summarized in Table~\ref{tab:config}, and their projected sensitivities are presented in Fig.~\ref{fig:reach}. We have conservatively restricted our analysis to frequencies above $1\,\mathrm{Hz}$, corresponding to axion masses above $m_a\simeq 4\times10^{-15}\,\mathrm{eV}$, to mitigate potential contamination from relative gondola roll.

The key experimental question is whether the polarization stability required to achieve the projected sensitivity can be attained under stratospheric flight conditions. A ground-based demonstration of the optical link using the proposed polarimeters and turnaround optics would constitute a natural first step, enabling direct characterization of the noise spectra of the apparatus. A subsequent Pathfinder flight would establish the polarization noise under realistic operating conditions, quantify the impact of gondola motion, and assess the feasibility of extending the search to lower axion masses.
The proposed architecture also offers several possibilities for further development. Depending on the measured noise properties, sensitivity could be improved through longer baselines, multiple passes between two balloons, or the simultaneous operation of several independent balloon pairs.
Beyond oscillating axion dark matter, the same experimental architecture could be adapted to search for transient polarization rotations associated with axion domain-wall crossings, complementing spin-based searches~\cite{Pospelov:2012mt,GNOME:2023rpz}. Moreover, the development of long-distance, polarization-preserving optical links in the stratosphere is of independent interest for quantum communication~\cite{Raja_baloon,QKDturb}.
KEBBABS thus opens a complementary route to axion dark matter searches, exploiting the flexibility and scalability of stratospheric platforms to access optical baselines beyond those achievable in conventional laboratory experiments. 

\begin{acknowledgments}
We thank Diego Blas, Valerie Domcke, Itay Bloch, Mario Reig, and Ippei Obata for their helpful comments on the manuscript. 
%We also acknowledge the use of AI tools developed by Anthropic and OpenAI to search the literature, explore and refine the experimental concept, and review the manuscript for clarity.

\end{acknowledgments}

\appendix

% ============================================================
\section{Signal derivation and response}\label{app:signal}
% ============================================================

\subsection{Axion Birefringence}
\label{app:endpoint}

Consider the propagation of the electromagnetic field $A^\mu$ in an axion
background. The relevant Lagrangian density is
\begin{equation}
\mathcal{L}
=-\frac{1}{2}\,\partial_\mu a\,\partial^\mu a-V(a)
-\frac{1}{4}\,F_{\mu\nu}F^{\mu\nu}
-\frac{g_{a\gamma}}{4}\,a\,F_{\mu\nu}\tilde{F}^{\mu\nu},
\label{eq:lagrangian-app}
\end{equation}
where $\tilde{F}^{\mu\nu}=\tfrac{1}{2}\epsilon^{\mu\nu\rho\sigma}F_{\rho\sigma}$
and, for the dark matter background of interest here,
$V(a)=\tfrac{1}{2}m_a^2a^2$.

We work in Coulomb gauge, $\vec{\nabla}\cdot\vec{A}=0$, and treat the axion as
spatially homogeneous over the apparatus: at the masses considered here its de
Broglie wavelength exceeds the baseline by many orders of magnitude, so
$a=a(t)$ throughout. Decomposing the transverse field on the circular
polarization basis $\hat{e}_{L/R}=(\hat{e}_1\pm i\,\hat{e}_2)/\sqrt{2}$, the two
helicity modes decouple and each obeys
\begin{equation}
\ddot{A}_{L/R}(t,k)
+k^2\left(1\mp\frac{g_{a\gamma}\dot{a}}{k}\right)A_{L/R}(t,k)=0,
\label{eq:helicity-eom}
\end{equation}
with the upper sign for $L$ and the lower for $R$, and a dot denoting $d/dt$.

The axion background therefore makes the two circular polarizations travel at
slightly different speeds,
\begin{equation}
c_{L/R}\simeq1\mp\frac{g_{a\gamma}\dot{a}}{2k},
\end{equation}
recovering Eq.~\eqref{eq:phasevelocity} of the main text. Equivalently, at
fixed optical frequency $\omega$ and to first order in $g_{a\gamma}$, the two
wavenumbers are
\begin{equation}
k_{L/R}=\frac{\omega}{c_{L/R}}\simeq\omega\pm\frac{g_{a\gamma}\dot{a}}{2}.
\label{eq:wavenumbers}
\end{equation}
This first-order treatment assumes the axion oscillation to be slow compared
with the optical period, $\omega_a\ll\omega$, which is amply satisfied for the
mass range considered.

The two helicities accumulate a relative phase along the path. Along the
unperturbed ray $dz=c\,dt$, so the phase difference collected between emission
at $t_0$ and detection at $t_1$ is
\begin{equation}
\Delta\varphi=\int_{t_0}^{t_1}\!\left(k_L-k_R\right)c\,dt
=g_{a\gamma}\int_{t_0}^{t_1}\!\dot{a}\,dt
=g_{a\gamma}\left[a(t_1)-a(t_0)\right].
\label{eq:phase-difference}
\end{equation}
The integral telescopes: whatever the axion field does along the way drops out,
and only its values at the two ends of the trajectory remain. A linear
polarization is an equal superposition of the two circular ones, and a relative
circular phase $\Delta\varphi$ corresponds to a rotation of the linear polarization direction by $\theta=\Delta\varphi/2$, which yields
Eq.~\eqref{eq:endpoint}.

\subsection{Even and odd reflection numbers}\label{app:parity}

Each mirror reflection reverses the photon helicity, exchanging left- and right-circular polarization. Since the sign of the axion-induced rotation depends on the helicity, the number of reflections at the remote terminal must be even to avoid a partial cancellation of the signal along the return path.

Indeed, with an even number of reflections, the helicity is restored before the return path, so the rotations accumulated along the two legs add. The resulting signal is equivalent to that of a single propagation over a distance \(2L\), with response
\[
F_{\rm even}(m_a)=|\sin x|,\qquad x=\frac{\omega_a L}{c}.
\]
With an odd number of reflections, the helicity is reversed and the contributions from the two legs partially cancel, giving
\[
F_{\rm odd}(m_a)=2\sin^2(x/2).
\]
This distinction is particularly important at low axion masses (\(x\ll1\)), where \(F_{\rm even}\simeq x\), while \(F_{\rm odd}\simeq x^2/2\). An even number of reflections is therefore essential to avoid the stronger low-mass suppression.
\section{Propagation channel}\label{app:link}
% ============================================================

This appendix describes how much power returns to the polarimeter after the
round trip, how the geometry and the motion of the turnaround affect it, and
how the atmosphere acts on the beam and on its polarization. It covers the
received laser power (\ref{app:power}), with its geometric
(\ref{app:geo}) and optical (\ref{app:etafixed}) contributions, extinction
(\ref{app:ext}), turbulence (\ref{app:turb}), scattering-induced
depolarization (\ref{app:depol}), and magneto-optical rotation
(\ref{app:faraday}).

%===========================================================================
\subsection{Received laser power}\label{app:power}
%===========================================================================

We compute the laser power that reaches the polarimeter after the round trip
as
\begin{equation}
 P_{\rm rx}=P_{\rm tx}\,\eta_{\rm geo}\,\eta_{\rm opt},
 \label{eq:prx-link}
\end{equation}
where $P_{\rm tx}$ is the laser output power, i.e.\ the power entering the
transmit telescope. The losses are split into two contributions. The
geometric factor $\eta_{\rm geo}$ accounts for the finite size of the three
apertures, for diffraction, and for the pointing errors of the two terminals. The
optical factor $\eta_{\rm opt}$ accounts for the internal losses of the
optics, namely the telescopes, the turnaround mirrors, and the polarimetric
components. Atmospheric extinction is negligible at float
altitude, as we discuss in App.~\ref{app:ext}.
For the optimized parameters of Table~\ref{tab:config}, using Eq. \ref{eq:prx-link} the received power is
\begin{align}
 P_{\rm rx}^{\rm PF}&=2.74\times10^{-3}~{\rm W},\notag\\
 P_{\rm rx}^{\rm FT}&=0.14~{\rm W},\notag\\
 P_{\rm rx}^{\rm FUT}&=10.5~{\rm W}.
 \label{eq:prxvalues}
\end{align}

%===========================================================================
\subsubsection{Geometric losses}\label{app:geo}
%===========================================================================

The geometric factor depends on diffraction, pointing fluctuations, and
the finite size of the optical apertures. To model these effects, we
approximate the propagating beam as Gaussian\footnote{Further optimization
of the initial laser beam profile, including the use of non Gaussian beams,
as well as more sophisticated collimation and beam shaping techniques, could
increase the received power and reduce the optical aperture requirements.}.

The laser has a fixed input beam radius $w_{\rm laser}$.\footnote{We define
$w$ as the radius at which the intensity falls to $1/e^2$ of its peak
value, i.e., the radius enclosing approximately $86.5\%$ of the total
optical power.} The transmitting telescope expands and collimates the beam
to reduce its diffraction-induced divergence. Denoting the optical
magnification of the telescope by $\mathcal{M}$, the gaussian beam radius at the exit of the
transmitting telescope is $w_0=\mathcal{M}\,w_{\rm laser}$. It's radius after
propagation over a distance $z$ is
\begin{equation}
 w_G(z)=w_0\sqrt{1+\left(\frac{z}{z_R}\right)^2},\qquad
 z_R=\frac{\pi w_0^2}{\lambda}.
 \label{eq:gaussian-propagation}
\end{equation}
We can seen from Eq.~\eqref{eq:gaussian-propagation} that for a fixed
distance $z$ there is an optimal initial radius $w_0^\textit{opt}$ that minimizes
$w_G(z)$, and hence maximizes the received power at the terminal at distance z.  Writing
$w_G^2(z)=w_0^2+(\lambda z/\pi w_0)^2$ and minimizing $w_G(z)$ with respect to $w_0$
gives
\begin{equation}
 w_0^{\rm opt}=\sqrt{\frac{\lambda z}{\pi}},\qquad
 w_G^{\min}(z)=\sqrt{\frac{2\lambda z}{\pi}}.
 \label{eq:optimal-waist}
\end{equation}
An initial radius $w_0 < w_0^\mathrm{opt}$ leads to a larger divergence and hence to a larger $w_G(z)$. Conversely, an initial radius $w_0 > w_0^\mathrm{opt}$ has a smaller divergence, but since the beam starts from a larger radius, at distance $z$ it still covers a larger area than $w_G^\mathrm{min}$.
In all three technological levels we assume a laser with beam radius $w_\mathrm{laser} = 1\,\mathrm{mm}$. Knowing $w_0^\mathrm{opt}$ for a fixed $z$, we can then find the optical magnification required to expand the beam to $w_0^\mathrm{opt}$, namely
\begin{equation}
  \mathcal{M}^\mathrm{opt} = \frac{w_0^\mathrm{opt}}{w_\mathrm{laser}}
  = \frac{1}{w_\mathrm{laser}}\sqrt{\frac{\lambda z}{\pi}}.
  \label{eq:optimal-magnification}
\end{equation}
Once the magnification $\mathcal{M}$ is fixed, we determine the diameter $D_\mathrm{T}$ of the transmitting telescope by requiring that its aperture intercepts almost all the power of the expanded Gaussian beam after the magnification, while not exceeding the maximum aperture $D_{\max}$ allowed at each technological level. 
In general, a circular aperture of diameter $D$ collects the following fraction of the
power of a centred Gaussian beam of radius $w$:
\begin{equation}
 \eta(D,w)=1-\exp\left[-\frac{D^2}{2w^2}\right].
 \label{eq:etaleg}
\end{equation}
We require the transmitting aperture to intercept nearly all the power of the
expanded beam. Specifically, we impose a collected fraction of $99\%$. Setting
$\eta(D_{\rm T},w_0)=0.99$ gives
\begin{equation}
 D_{\rm T}
 \simeq 3.03\,w_0
 =3.03\,\mathcal{M}\,w_{\rm laser}.
 \label{eq:transmit-diameter}
\end{equation}
In the design optimization of Appendix~\ref{app:optimization} the waist $w_0$
is a free parameter, bounded by the aperture limit,
\begin{equation}
 w_0\leq\frac{D_{\max}}{3.03},
 \qquad
 \mathcal{M}=\frac{w_0}{w_{\rm laser}},
 \label{eq:w0-bounded}
\end{equation}
and $D_{\rm T}$ follows from Eq.~\eqref{eq:transmit-diameter}.

Residual transmitter pointing fluctuations produce beam wander. Assuming
independent Gaussian angular fluctuations with RMS $\sigma_p$ per transverse
axis, we approximate the outgoing intensity by a Gaussian
of effective radius
\begin{equation}
 w_{\rm out}(z)=\sqrt{w_G^2(z)+4z^2\sigma_p^2},
 \label{eq:weff}
\end{equation}
where $\sigma_p$ is given in Table~\ref{tab:config}. A transmitter pointing error $\sigma_p$, acting over the full path, produces
an RMS transverse displacement $\sigma_p\,(2L)$ at the receiving terminal. A
pointing error $\varepsilon_{\rm turn}$ of the turnaround deflects the
returning beam by $2\varepsilon_{\rm turn}$, but only over the return path
$L$, producing an RMS displacement $(2\varepsilon_{\rm turn})\,L$. The two
errors therefore enter with the same weight. Since the two displacements are
independent, they add in quadrature. As in Eq.~\eqref{eq:weff}. The effective
radius of the returning beam is therefore
\begin{equation}
 w_{\rm ret}(2L)=
 \sqrt{w_G^2(2L)+16L^2\sigma_p^2+16L^2\varepsilon_{\rm turn}^2}.
 \label{eq:wret}
\end{equation}
This expression is conservative. Owing to its geometry, the roof mirror
automatically cancels its own pointing errors along the vertical axis (for
the orientation shown in Fig.~\ref{fig:setup}), while along the horizontal
axis it behaves as a flat mirror. The turnaround error
$\varepsilon_{\rm turn}$ therefore acts only along the horizontal axis, and
applying it to both axes, as in Eq.~\eqref{eq:wret}, overestimates the
broadening of the returning beam.
The flat-mirror turnaround redirects the beam but, due to its finite
size, intercepts only a fraction of the incident power. The power
ultimately collected also depends on the size of the final receiving telescope.
We therefore use Eq.~\eqref{eq:etaleg} to estimate the fractions of
power intercepted by the turnaround and collected by the receiving
telescope. Including the $99\%$ transmission of the transmit aperture,
the geometric factor is
\begin{equation}
 \eta_{\rm geo}=0.99\
 \eta\!\left(D_{\rm turn},w_{\rm out}(L)\right)\,
 \eta\!\left(D_{\rm R},w_{\rm ret}(2L)\right).
 \label{eq:etageo}
\end{equation}

%===========================================================================
\subsubsection{Optical losses}\label{app:etafixed}
%===========================================================================
The optical factor collects the losses of the optics along the path, which do
not depend on the link geometry. We write it as
\begin{equation}
 \eta_{\rm opt}=\eta_{\rm tel}^{2}\,R_{\rm turn}^{2}\,\eta_{\rm pol}.
 \label{eq:etafixed}
\end{equation}

The factor $\eta_{\rm tel}$ is the transmission we assume for each telescope,
taken equal for the transmitting and the receiving one. 
The factor $R_{\rm turn}^{2}$ accounts for the reflectivity of the two
turnaround mirrors, designed for operation at $45^\circ$.
Finally, $\eta_{\rm pol}$ accounts for the losses of the polarization
measurement setup at the active terminal. Table~\ref{tab:etafixed} lists the
values we adopt.

\begin{table}[t]
\centering
\footnotesize
\begin{tabular}{@{}lcc@{}}
\toprule
Component & PF, FT & FUT \\
\midrule
Telescope, per pass $\eta_{\rm tel}$ & 0.95 & 0.98 \\
Turnaround (2 mirrors) $R_{\rm turn}^2$ & 0.99 & 0.99 \\
Polarimetric optics $\eta_{\rm pol}$ & 0.96 & 0.97 \\
\midrule
Optical factor $\eta_{\rm opt}$ & 0.86 & 0.92 \\
\bottomrule
\end{tabular}
\caption{The optical efficiencies are assumed values guided by manufacturer specifications for commercially available near-infrared telescopes and optical components. The Futuristic configuration adopts more optimistic efficiencies, reflecting higher performance optics.}
\label{tab:etafixed}
\end{table}

%%%%%%%%%%%%%%%%%%%%%%%%%%%%
\subsection{Atmospheric effects along the beam path}\label{app:atm}\label{app:ext}\label{app:turb}\label{app:depol}

We assume both balloons float at $h_{\rm float}=40$ km. Because the line of
sight between them is a chord, the beam passes below the float altitude, with
minimum altitude at mid-path
\begin{equation}
 h_{\rm min}\simeq h_{\rm float}-\frac{L^2}{8R_\oplus},
\end{equation}
where $L$ is the one-way balloon separation and $R_\oplus\simeq6400$ km. For
the longest baseline considered, $L=600$ km, this gives $h_{\rm min}\simeq33$
km. More generally, the beam stays above $30$ km as long
as
\begin{equation}
 L\lesssim\sqrt{8R_\oplus\left(h_{\rm float}-30~{\rm km}\right)}
 \simeq7.1\times10^2~{\rm km}.
\end{equation}
In the following we therefore evaluate all atmospheric effects at $30$ km.
This is a conservative choice: most of the path lies well above this altitude,
especially for the shorter baselines.

At $1550$ nm, extinction is produced mainly by molecular scattering, aerosols,
and absorption. Each contribution is described by an extinction coefficient
$\beta$, so that a path of length
$L_{\rm path}$ transmits a fraction $e^{-\beta L_{\rm path}}$. Molecular
scattering is proportional to the air density. At sea level its extinction
coefficient is $\beta_{\rm R}\simeq1.8\times10^{-4}~{\rm km^{-1}}$~\cite{Bucholtz1995};
since the density at $30$ km is about $70$ times
smaller~\cite{USStandard1976}, it drops to
$\beta_{\rm R}\simeq3\times10^{-6}~{\rm km^{-1}}$. The main stratospheric
aerosol layer lies near $20$--$25$ km, and near-infrared aerosol extinction
around $30$--$40$ km is below ${\sim}10^{-5}~{\rm km^{-1}}$~\cite{Kremser2016}.
Absorption is subdominant, since $1550$ nm lies in a low-loss atmospheric
window. We therefore adopt a conservative total extinction coefficient
$\beta=1.5\times10^{-5}~{\rm km^{-1}}$. Even in
this worst case, applying the same coefficient thought all the path, the atmospheric transmission exceeds $0.98$ for all
configurations. In practice the losses will be even smaller
since, as already said,  most of the beam travels at altitudes well above $30$ km, where both the
air density and the aerosol content are lower. We therefore neglect
atmospheric extinction in the link budget.

Turbulence does not affect the polarization but it could temporary deflect the laser beam, thus contributing to the pointing error. Above ${\sim}15$ km it occurs
only in thin, intermittent layers, and measurements up to $30$ km show
essentially no turbulence outside them~\cite{Schneider2017,Haack2014}.
Its main effect would be a random angular deflection of the beam, known as
turbulent beam wander. For a collimated Gaussian beam of initial radius $w_0$
propagating over a distance $L$ through turbulence of uniform strength, the
RMS wander angle is~\cite{AndrewsPhillips2005}
\begin{equation}
 \theta_{\rm w}\simeq\sqrt{2.42\,C_n^2\,L\,w_0^{-1/3}},
 \label{eq:wander}
\end{equation}
where $C_n^2$ is the refractive index structure constant, which measures the
strength of the refractive index fluctuations. The value of $C_n^2$ at a given altitude can be estimated with the
Hufnagel-Valley model~\cite{Valley1980,AndrewsPhillips2005}, which gives
$C_n^2\simeq2\times10^{-21}\,\mathrm{m^{-2/3}}$ at $30$ km and smaller
values above.Taking this value along the whole path,
Eq.~\eqref{eq:wander} gives $\theta_{\rm w}\simeq0.02$, $0.03$ and
$0.06\,\mu$rad for the three configurations, about an order of magnitude or
more below the pointing errors of Table~\ref{tab:config}.

Finally, scattering does not degrade the polarization measurement. Photons
scattered once are removed from the beam and only contribute to the
extinction discussed above. A depolarized contribution requires a photon to be
scattered out of the beam and then scattered back into it, with a probability
of order $\tau^2$, where $\tau=\beta L_{\rm path}\lesssim2\times10^{-2}$ is the
optical depth along the longest path. The depolarized fraction is therefore
below $4\times10^{-4}$, a conservative bound since only a tiny fraction of
twice-scattered photons is redirected along the beam. Such a small, static
depolarization only marginally reduces the polarization contrast and cannot
mimic a rotation oscillating at the axion frequency; its effect is therefore
negligible.

\subsection{Faraday rotation}\label{app:faraday}
At the laser angular frequency $\omega$, the geomagnetic field
$\mathbf B_0$ induces a magneto-optical response in air~\cite{Sigwarth2022}.
The corresponding cycle-averaged interaction is
\begin{equation}
\overline{\mathcal L}_F
=\frac{i\epsilon_0\gamma(\omega)}{4}\,
\mathbf B_0\cdot(\mathbf E_\omega^*\times\mathbf E_\omega),
\end{equation}
where $\mathbf E_\omega$ is the complex electric-field amplitude and
$\gamma(\omega)$ is the magneto-optical coefficient. For a mode of
helicity $h=\pm1$ propagating along $\hat{\mathbf k}$,
$\mathbf e_h^*\times\mathbf e_h=ih\,\hat{\mathbf k}$. The two helicities
therefore acquire opposite wavenumber shifts,
$\delta k_h=-hV\,\mathbf B_0\cdot\hat{\mathbf k}$, where
$V=\omega\gamma/(2nc)$ is the Verdet constant and $n$ is the refractive
index. Their relative phase rotates the plane of linear polarization.

At $1550\,\mathrm{nm}$, neutral air dominates the one-way rotation
because the stratospheric free-electron density is very low~\cite{LehtinenInan2007}.
In the dilute-gas limit, we take $V$ to scale with molecular number
density~\cite{Gianella2019}. For fixed air composition, this gives
\begin{equation}
\phi_{\rm air}^{(1)}
\simeq V_0\int_A^B
\frac{\rho_{\rm air}(s)}{\rho_0}\,B_\parallel(s)\,ds ,
\end{equation}
where $B_\parallel=\mathbf B_0\cdot\hat{\mathbf k}$ on the outgoing
path and $\rho_0$ is the sea-level air density. Laboratory measurements
near $1550\,\mathrm{nm}$ give
$V_0\simeq(1.2$--$2.8)\times10^{-4}\,
\mathrm{rad\,T^{-1}\,m^{-1}}$~\cite{Phelps2015,Guo2024}.
For the $600\,\mathrm{km}$ baseline, the beam altitude ranges from
$40\,\mathrm{km}$ at the terminals to $33\,\mathrm{km}$ at the midpoint.
Integrating the standard atmospheric density along this path~\cite{USStandard1976}
gives $|\phi_{\rm air}^{(1)}|\simeq25$--$60\,\mu\mathrm{rad}$ for a
constant $|B_\parallel|=50\,\mu\mathrm{T}$.

All three KEBBABS configurations measure a round trip. The ideal
two-mirror turnaround preserves helicity and reverses
$\hat{\mathbf k}$, so $\delta k_h$ changes sign for each helicity.
If the beam retraces its path and the medium and field remain stationary,
the Faraday phases cancel exactly~\cite{Martinelli2016}. Differences
between the two paths or changes during the round trip can leave a
residual which is stationary in the frequency range considered. Variations on minute-to-hour timescales fall below the
$1\,\mathrm{Hz}$--$5\,\mathrm{kHz}$ axion search band.

% ============================================================
\section{Polarimetric readout}\label{app:polarimeter}\label{app:polarimetric-readout}
% ============================================================

The polarimeter converts a tiny polarization rotation into a measurable
power imbalance between two photodetectors. This appendix derives the
two-port estimator used to reconstruct the polarization rotation. Photon
shot noise and the propagation of technical disturbances are treated
separately in Sec.~\ref{sec:noise} and Appendix~\ref{app:tech}.

The architecture is the two-port readout of DANCE
Act-1~\cite{Oshima2023}. A half-wave plate (HWP) rotates the polarization by
a small, calibrated reference angle; a polarizing analyzer then splits the
beam into its dominant $s$ component and its weak $p$ component, whose
powers are measured simultaneously by two photodiodes. No modulation or
lock-in demodulation is required, and normalizing by the total received
power makes the estimate immune to fluctuations of the laser power and of
the link transmission that are common to the two ports.

Let $\theta(t)$ be the polarization rotation accumulated along the
optical link. Immediately before the HWP, the Jones vector of the returning
beam is
\begin{equation}
 \mathbf E_{\rm ret}(t)=\sqrt{P_{\rm rx}(t)}
 \begin{pmatrix}
 \cos\theta(t)\\
 \sin\theta(t)
 \end{pmatrix},
\end{equation}
where the first and second components denote the $s$ and $p$ axes of the
analyzer. The HWP adds the known reference offset
\begin{equation}
 \epsilon=2\theta_{\rm HWP}
\end{equation}
to the polarization angle, so the field incident on the analyzer is
\begin{equation}
 \mathbf E_{\rm ana}(t)=\sqrt{P_{\rm rx}(t)}
 \begin{pmatrix}
 \cos\alpha(t)\\
 \sin\alpha(t)
 \end{pmatrix},
 \qquad
 \alpha(t)=\epsilon+\theta(t).
\end{equation}
The ideal analyzer outputs are
\begin{align}
 P_s(t)&=P_{\rm rx}(t)\cos^2\alpha(t),\notag\\
 P_p(t)&=P_{\rm rx}(t)\sin^2\alpha(t),
 \label{eq:polarimeter-powers}
\end{align}
so that $P_s+P_p=P_{\rm rx}$: measuring both ports gives the angle and the
total power at once. The exact estimator of the rotation is
\begin{equation}
 \widehat{\theta}(t)=
 \arcsin\!\sqrt{\frac{P_p(t)}{P_s(t)+P_p(t)}}-\epsilon.
 \label{eq:exact-estimator}
\end{equation}
For $|\epsilon|,|\theta|\ll1$, this reduces to the expression used in
the main text and in DANCE Act-1,
\begin{equation}
 \widehat{\theta}(t)\simeq
 \sqrt{\frac{P_p(t)}{P_s(t)+P_p(t)}}-\epsilon.
 \label{eq:small-angle-estimator}
\end{equation}
Equivalently,
\begin{equation}
 P_p(t)\simeq P_{\rm rx}(t)
 \left[\epsilon^2+2\epsilon\,\theta(t)\right],
\end{equation}
which shows explicitly the role of the HWP: it converts the rotation into a
power signal that is linear, rather than quadratic, in $\theta$. A
static error in $\epsilon$ only shifts the mean reconstructed angle and is
irrelevant for a search of an oscillating signal, whereas fluctuations of
$\epsilon$ within the analysis band mimic the signal directly and are a
technical-noise source (Appendix~\ref{app:tech}).
\section{Technical noise}\label{app:tech}
% ============================================================
% REV5: rewritten. Values rounded to 5e-9, 5e-10, 5e-11 rad/sqrt(Hz);
% REV5: Table I and the three reach values must be recomputed (effect <1%).
% REV5: new bib entries needed: He2016, Li2014, AutoBalanced2011,
% REV5: SuperBIT2024roll (Voyer et al. 2026, arXiv:2607.28816), WASP.

The technical term $S_{\theta,\rm tech}$ in Eq.~\eqref{eq:line-noise}
collects every instrumental effect that mimics a polarization rotation in
the analysis band, $1\,\mathrm{Hz}$--$5\,\mathrm{kHz}$. We anchor the adopted
values to demonstrated polarimetric performance and treat the effects that
are specific to the balloon platform as requirements, which we then compare
with demonstrated gondola performance. The two port readout of
Sec.~\ref{sec:readout} is one way to meet these requirements, not a
constraint of the concept: any polarimetric scheme reaching the same
rotation ASD would serve equally well.

\subsection{Demonstrated polarimetric floors}\label{app:tech-demo}

Without any optical cavity, modulation free polarimeters based on a
polarizer-analyzer pair and two photodetectors\footnote{Polarimeters
of the PVLAS type instead impose a known, fast oscillation on the
polarization and extract the tiny rotation as a shift of that oscillation,
rather than reading it directly~\cite{PVLAS}. This moves the measurement
away from the low-frequency noise of laser and electronics. The cost is an
additional active element whose own mechanical and thermal fluctuations
produce spurious signals at that frequency and must be controlled to the
same level as the effect itself. A similar scheme, although applied to
microwave rather than optical polarization, has already flown on a
stratospheric balloon in the EBEX CMB polarimeter~\cite{EBEX2018}. For
simplicity, we restrict ourselves to the modulation-free scheme.\label{fn:modulated}} have
reached $7$--$9\times10^{-9}\,\mathrm{rad}/\sqrt{\mathrm{Hz}}$ at the
shot-noise limit with mW powers~\cite{He2016,Li2014}, in one case with a
noise spectrum that remains flat at the shot-noise level down to
$1\,\mathrm{Hz}$, the lower edge of our analysis band~\cite{He2016}, and
about $3\times10^{-8}\,\mathrm{rad}/\sqrt{\mathrm{Hz}}$ with a commercial
balanced receiver~\cite{Chang2011}. These measurements suggest that, in the
frequency band of our proposal, the intrinsic floor of a modulation free
polarimeter lies at or below the shot noise.

The two port readout adopted in this work, as already described, is the same used in DANCE Act-1~\cite{Oshima2023},
which reached $3.4\times10^{-6}\,\mathrm{rad}/\sqrt{\mathrm{Hz}}$ at
$5\,\mathrm{Hz}$ with $150\,\mathrm{mW}$ of transmitted power, five orders
of magnitude above its shot noise. This figure is an upper bound rather
than a floor for our purposes, the limiting mechanisms identified by the
DANCE collaboration, laser frequency noise and cavity vibrations converted
into apparent rotation by the MHz scale splitting between the $s$ and $p$
cavity resonances, are intrinsic to the resonant cavity and have no
counterpart in a single pass free space link. The same readout without the
cavity is expected to behave as the single pass polarimeters quoted above.

\subsection{Adopted values}\label{app:tech-values}

We require the technical floor of each configuration not to exceed its
shot noise, so that the projected sensitivity is shot noise limited by
construction, and we adopt
\begin{equation}
S_{\theta,\rm tech}^{1/2}=
5\times10^{-9},\quad 5\times10^{-10},\quad 5\times10^{-11}
\ \mathrm{rad}/\sqrt{\mathrm{Hz}}
\label{eq:techvalues}
\end{equation}
for the Pathfinder, Feasible today and Futuristic configurations,
respectively, taken flat across the analysis band and referred to a single
round trip line. The three values are motivated as follows:

\begin{itemize}
\item \emph{Pathfinder.} With $P_{\rm rx}\simeq2.7\,\mathrm{mW}$ the shot
noise is $5.4\times10^{-9}\,\mathrm{rad}/\sqrt{\mathrm{Hz}}$, and
shot-noise-limited operation at this power and down to $1\,\mathrm{Hz}$ has
been demonstrated~\cite{He2016,Li2014}. The adopted value is therefore
supported by existing measurements.
\item \emph{Feasible today.} With $P_{\rm rx}\simeq0.14\,\mathrm{W}$ the
shot noise drops to $7.5\times10^{-10}\,\mathrm{rad}/\sqrt{\mathrm{Hz}}$, and we
adopt a technical floor of $5\times10^{-10}\,\mathrm{rad}/\sqrt{\mathrm{Hz}}$,
slightly below it. Existing 
measurements reached $7\times10^{-9}\,\mathrm{rad}/\sqrt{\mathrm{Hz}}$, but
were limited by shot noise at mW powers~\cite{He2016,Li2014}, so they do not
yet tell us how far below this level the technical floor lies. Increasing
the power from mW to $\sim 0.1\,\mathrm{W}$ does not by itself introduce new
noise, the angle-equivalent electronic noise decreases faster than the shot noise~\footnote{At fixed bias $\epsilon$, a small rotation produces a
weak-port power change $\delta P_p \simeq 2\epsilon P_{\rm rx}\,\delta\theta$.
In the linear regime and at fixed gain, the electronic current noise
is approximately independent of $P_{\rm rx}$, whereas the shot-noise
current amplitude scales as $\sqrt{P_{\rm rx}}$. Dividing by the
angular response therefore gives noise amplitudes in
$\mathrm{rad}/\sqrt{\mathrm{Hz}}$ that scale as $P_{\rm rx}^{-1}$ and
$P_{\rm rx}^{-1/2}$, respectively.},
and laser intensity noise is rejected by the two port normalization
independently of power. The ground link test and the
Pathfinder flight will establish whether the setup remains
shot noise limited at higher power or, if not, which technical sources
dominate and how the setup must be improved to suppress them.
\item \emph{Futuristic.} With $P_{\rm rx}\simeq10.5\,\mathrm{W}$ the shot
noise is $8.7\times10^{-11}\,\mathrm{rad}/\sqrt{\mathrm{Hz}}$, and the
adopted technical floor is $5\times10^{-11}\,\mathrm{rad}/\sqrt{\mathrm{Hz}}$,
one order of magnitude below the Feasible today value. As discussed above,
existing measurements do not constrain the technical floor at this level,
and demonstrating it remains a development goal. The high power does not by itself
complicate the detection, since the $p$ port receives only
$\epsilon^{2}P_{\rm rx}\simeq12\,\mathrm{mW}$ for $\epsilon=3\times10^{-2}$,
and the $s$ port, used only for normalization, can be attenuated. The main challenge is instead keeping the readout and link optics stable
at the required level. Should the ground tests and the
Pathfinder flight reveal noise sources that the two port scheme cannot
suppress, the readout can be adapted to the dominant noise, for example 
with a modulated scheme (see footnote~\ref{fn:modulated}).
\end{itemize}

\subsection{Roll of the gondolas}\label{app:tech-roll}\label{app:optical-train}
Rotations of the gondolas about the two axes transverse to the line of sight
(tip and tilt) deflect the beam and are corrected by the pointing system, their
residual only modulates the received power (Appendix~\ref{app:geo}). A
relative rotation of the two gondolas about the line of sight (roll), instead,
is not corrected and directly produces a polarization rotation.

The roll motion of a balloon gondola is expected to lie mostly below the
frequency band shown in Fig.~\ref{fig:reach}. It is dominated by the
pendulation of the flight train: the in flight attitude spectra of SuperBIT
show a dominant pendulation peak near $0.5\,\mathrm{Hz}$, with higher order
modes below $1\,\mathrm{Hz}$~\cite{SuperBIT2024}. This motivates the
$1\,\mathrm{Hz}$ lower edge of the band.

Residual roll within the band can be reduced by passive isolation or active
stabilization of the turnaround and polarimeter mounts, and measured with
gyroscopes on the supports of the two gondolas. Since the induced polarization
rotation is linear in the relative roll, its contribution can then be
subtracted from the polarization data. An
optical tracking system could additionally provide an independent, monitor of the relative orientation of the two gondolas.

\subsection{Balloon separation}\label{app:separation}

Stratospheric balloons drift with the winds at float altitude. Their
horizontal position can nevertheless be controlled, for example by changing
altitude to reach winds blowing in different directions, as done by Project
Loon, or with lateral propellers~\cite{Loon, LoonLibrary}.
Here we motivate the fluctuations of the balloon separation around its target
value assumed in the main text.

\begin{figure}[t]
  \centering
  \includegraphics[width=\columnwidth]{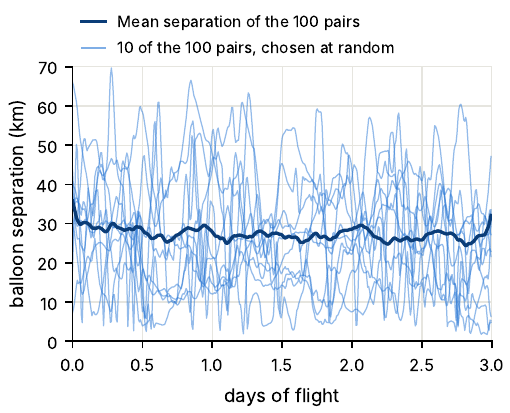}
  \caption{Separation between pairs of Project Loon balloons at float, from
  the flight data of Ref.~\cite{LoonData} (October to December 2019). Among
  the pairs with a three day mean separation between $20$ and $40$ km, we
  select the $100$ whose separation was most stable over three days. Thin
  lines show $10$ of them chosen at random, and the thick line shows the mean
  separation of all $100$. The separation of each pair fluctuates around its
  three day mean with a standard deviation of $11.5$ km on average.}
  \label{fig:loon}
\end{figure}

For the Pathfinder the target is $L\simeq30$ km.
Figure~\ref{fig:loon} shows the mean L between pairs of balloons in the
flight data of Project Loon~\cite{LoonData}, recorded in one of their campaigns between October and
December 2019. Their mean
separation is about $27$ km, close to the Pathfinder target, and it fluctuates around its three day mean with a standard
deviation of $11.5$ km on average. The value $\sigma_{\rm rel}=10$ km adopted
for the Pathfinder is therefore close to what Loon achieved. Most importantly, Loon
balloons were not steered to stabilize their relative distance, but to remain
over the areas they had to serve. These data thus describe balloons whose
separation is left free, and a two balloon mission steering towards a common
relative target is expected to keep the separation more stable.

For the Feasible today and Futuristic configurations the balloons are much
farther apart and may be carried by different air masses. We therefore assume
that each balloon is independently kept close to its own ground station, as
done by Project Loon in a $39$ day mission over the Pacific
Ocean~\cite{Bellemare2020}, where the balloons remained within $50$ km of
their station for $79\%$ of the time. We adopt a 
fluctuation of the separation with $\sigma_{\rm rel}=30$ km for the Feasible today and $\sigma_{\rm rel}=50$ km for the Futuristic.
% ============================================================

\section{Sensitivity calculation}\label{app:reach}
In this section, we make sense of the different scalings of Eq.~\eqref{eq:ampunc} and estimate the conversion between the nominal ${\rm SNR}=1$
sensitivities into the median expected
upper limits $95\%$.
We follow the statistical framework of
Refs.~\cite{Centers2021, Foster:2017hbq}, implemented in the DANCE
analysis~\cite{Oshima2023}, but introduce an equal-contribution
approximation suitable for a proposal-level estimate.

The Rayleigh distribution arises because the virialized axion field
is a superposition of many waves with random phases, the velocity spread of the
dark matter halo gives it a fractional linewidth of order $10^{-6}$, so the
oscillation stays phase-coherent only for about $Q\sim10^6$ cycles,
corresponding to a coherence time
$\tau_c\simeq Q/\nu_a$~\cite{Centers2021,Budker2014}. Over a time
short compared with the coherence time, the field can be written as
\begin{equation}
    a(t)\simeq a_c\cos(\omega_a t)+a_s\sin(\omega_a t).
\end{equation}

Both coefficients are sums of many random contributions and are
therefore approximately independent, zero-mean Gaussian variables.
Their variances are
\begin{equation}
\langle a_c^2\rangle=\langle a_s^2\rangle
=\frac{a_{\rm DM}^2}{2},
\qquad
a_{\rm DM}\equiv\frac{\sqrt{2\rho_{\rm DM}}}{m_a},
\end{equation}
using natural units for the field normalization. The oscillation
amplitude is the length of this two-dimensional Gaussian vector,
$a_0=\sqrt{a_c^2+a_s^2}$, and hence follows the Rayleigh distribution
\begin{equation}
p(a_0)=\frac{2a_0}{a_{\rm DM}^2}
\exp\!\left(-\frac{a_0^2}{a_{\rm DM}^2}\right),
\qquad a_0\geq0,
\end{equation}
as in Eq.~(2) of Ref.~\cite{Centers2021}.
Thus $a_{\rm DM}=\sqrt{\langle a_0^2\rangle}$ is the
RMS oscillation amplitude across field realizations,
rather than a fixed amplitude realized in every observation.

For our optical response, define
\begin{equation}
\mathcal R(m_a,L)\equiv
a_{\rm DM}|F_L(m_a)|,
\end{equation}
then the RMS oscillation amplitude of the rotation is $g_{a\gamma}\mathcal R$. 
As the balloons drift, their one-way separation \(L(t)\) varies during the observation. For a search that tracks the measured separation in the signal template, we define an effective RMS response
\begin{equation}
\mathcal R_{\rm eff}^2(m_a)
=a_{\rm DM}^2\left\langle
\sin^2\!\left(\frac{\omega_a L(t)}{c}\right)
\right\rangle_{\rm obs},
\end{equation}
where the average is taken over times when the optical link is operating. For the forecast, we model the separation as Gaussian with mean \(\bar L\) and standard deviation \(\sigma_{\rm rel}\). This gives
\begin{equation}\label{Eq:E6}
{\mathcal R_{\rm eff}^2(m_a)=
\frac{a_{\rm DM}^2}{2}
\left[1-e^{-2(\omega_a\sigma_{\rm rel}/c)^2}
\cos\!\left(\frac{2\omega_a\bar L}{c}\right)\right].}
\end{equation}
Since the response is linear,
the cosine and sine coefficients of the rotation signal are also
Gaussian, each with variance $g_{a\gamma}^2\mathcal R^2/2$.
To forecasts the sensitivity, we approximate the observation by independent, equivalent stretches
of duration $\Delta t=\min(T,\tau_c)$ and number $N=T/\Delta t$.
For locally constant one-sided noise PSD,
$S_\theta(\nu_a)=\sigma_\theta^2$, the measured quadratures $X_j,Y_j$
have noise variance $\sigma_q^2\simeq S_\theta/\Delta t$.
Assuming stationary and gaussian noise, after averaging over the random axion field, the total variance
of each measured coefficient is therefore
\begin{equation}
V=\sigma_q^2+\frac{g_{a\gamma}^2\mathcal R^2}{2}
=\sigma_q^2(1+\eta),
\qquad
\eta\equiv
\frac{g_{a\gamma}^2\mathcal R^2}{2\sigma_q^2}.
\end{equation}
Here $\eta$ is simply the ratio of signal variance to noise
variance in either measured quadrature.

The unknown signal phase motivates combining both measured
coefficients through their squared magnitude. We define the
dimensionless measured power
\begin{equation}
P_j\equiv\frac{X_j^2+Y_j^2}{2\sigma_q^2},
\end{equation}
normalized so that its noise-only mean is unity.
Because $X_j$ and $Y_j$ are independent Gaussians with variance
$V$, their squared magnitude follows a chi-squared distribution
with two degrees of freedom. Equivalently,
\begin{equation}
p(P_j\mid\eta)=
\frac{1}{1+\eta}
\exp\!\left[-\frac{P_j}{1+\eta}\right].
\end{equation}
This is the stochastic single-contribution likelihood of
Eq.~(A3) in Ref.~\cite{Oshima2023}, with the identification
$P_j=\rho_j/2$ and $\eta=\lambda_j^2$.
For this analytic forecast, we treat the time segments as independent
and assume the same expected ratio $\eta$ of signal power to noise
power in each segment. We account for the finite width of the axion
line through the coherence time $\tau_c$. A frequency-domain analysis
that resolves the line should also use the halo velocity distribution
to predict the fraction of signal power in each frequency
bin~\cite{Foster:2017hbq}.

For $N$ independent contributions, the summed power
$U=\sum_{j=1}^{N}P_j$ is a sum of identically distributed
exponentials, so that
\begin{equation}
\frac{U}{1+\eta}\sim{\rm Gamma}(N,1).
\end{equation}
Under the noise-only hypothesis, $U$ has mean $N$ and standard
deviation $\sqrt{N}$; a signal increases its mean by $N\eta$.
The ratio of this excess to the noise-only standard deviation
is therefore $\sqrt{N}\eta$. Setting this power signal-to-noise
ratio to unity gives
\begin{equation}
g_1\equiv g_{{\rm SNR}=1}
=
\frac{\sqrt{2}\sigma_q}{\mathcal R}N^{-1/4}
\simeq
\frac{\sqrt{2S_\theta(\nu_a)/\tau_c}}{\mathcal R}
\left(\frac{\tau_c}{T}\right)^{1/4},
\end{equation}
where we have used $N\simeq T/\tau_c$. This reproduces the
nominal sensitivity adopted in Sec.~\ref{sec:reach} and explains the scaling with $T\gtrsim\tau_c$ of Eq.~\eqref{eq:ampunc}. 

To estimate median expected $95\%$ upper limits, let $q_p(N)$ denote
the $p$-quantile of ${\rm Gamma}(N,1)$. Following
Ref.~\cite{Oshima2023}, we use the median noise-only observation
$U_{\rm med}=q_{0.5}(N)$ and solve
$\Pr(U\leq U_{\rm med}\mid g_{95}^{\rm med})=0.05$. This yields
\begin{equation}
{
\frac{g_{95}^{\rm med}}{g_1}
=N^{1/4}
\left[\frac{q_{0.5}(N)}{q_{0.05}(N)}-1\right]^{1/2}.
}
\label{eq:app_confidence_conversion}
\end{equation}
The conversion is larger at low masses where there are fewer coherence intervals, while a $N\gg1$, the conversion factor approaches the constant
$\sqrt{\Phi^{-1}(0.95)}\simeq1.283$, where $\Phi$ is the
standard normal cumulative distribution function.
For reference, at $m_a=10^{-14}\,\mathrm{eV}$, $\tau_c\simeq4.8\,\mathrm{d}$:
for the Pathfinder, Feasible today and Futuristic observation times
of $30$, $50$ and $365\,\mathrm{d}$, the conversion factors are
approximately $1.680$, $1.565$ and $1.371$, respectively.

% ============================================================
\section{Design optimization and parameter trade-offs}
\label{app:optimization}
% ============================================================

For each technological level we fix the quantities that define the assumed
technology level (laser power, pointing, optical throughput, integration time and
technical noise, Table~\ref{tab:config}) and choose the baseline and the three
apertures that give the best sensitivity at the reference mass:
\begin{equation}
 \min_{L,D_{\rm T},D_{\rm R},D_{\rm turn}}
 g_{a\gamma}(m_0),
 \qquad m_0=10^{-14}\,\mathrm{eV}.
 \label{eq:optproblem}
\end{equation}
In the optimization we use the RMS baseline response of Eq.~\eqref{Eq:E6} and optimize over the
mean balloon separation $\bar L$, at which the received power is evaluated.

The only constraints are upper bounds on the baseline and on the apertures,
listed in Table~\ref{tab:optimization}. The apertures are limited by what a
balloon gondola can carry~\cite{SunriseIII,BLAST}. The baseline is limited to
the $100$ km inter-balloon range demonstrated by Project Loon~\cite{Loon} for
the Pathfinder, and
to $500$ km and $700$ km for the Feasible today and the Futuristic level, respectively. To find the optimum we evaluate
$g_{a\gamma}(m_0)$ on a regular grid covering the whole allowed range: the
baseline from $0.1$ km to its upper bound in steps of $0.1$ km, and each of the
three apertures from $0.01$ m to its upper bound in steps of $0.01$ m. The grid contains to $7.5\times10^{7}$, $1.7\times10^{9}$ and $2.0\times10^{10}$
combinations for the Pathfinder, Feasible today and Futuristic levels,
respectively. The values in Table~\ref{tab:config} are the best design on the
grid.

\begin{table}[t!]
\centering
\vspace{0.5cm}
\footnotesize
\begin{tabular}{@{}lccc@{}}
\toprule
 & PF & FT & FUT \\
\midrule
Maximum target $L$ (km) & $100$ & $500$ & $700$ \\
Maximum aperture (m) & $0.5$ & $1.3$ & $1.5$ \\
Separation drift $\sigma_{\rm rel}$ (km) & $10$ & $30$ & $50$ \\
\midrule
Optimal $L$ (km) & $27.5$ & $226.4$ & $667.0$ \\
Optimal $D_{\rm T}$ (m) & $0.39$ & $1.13$ & $1.50^{*}$ \\
Optimal $D_{\rm R}$, $D_{\rm turn}$ (m) & $0.50^{*}$ & $1.30^{*}$ & $1.50^{*}$ \\
Reach ($10^{-12}$ GeV$^{-1}$) & $27.8$ & $0.273$ & $0.00630$ \\
\bottomrule
\end{tabular}
\caption{Upper bounds and separation drift of the optimization problem
\eqref{eq:optproblem}, and optimized parameters. The reach includes the drift of
the balloon separation through the RMS response of Eq.~\eqref{Eq:E6}, with
$\sigma_{\rm rel}$ the RMS fluctuation of the separation (App.~D\,4). An asterisk
marks parameters at their upper bound.}
\label{tab:optimization}
\end{table}
\begin{figure*}[h!]
\centering
\includegraphics[width=0.9\textwidth]{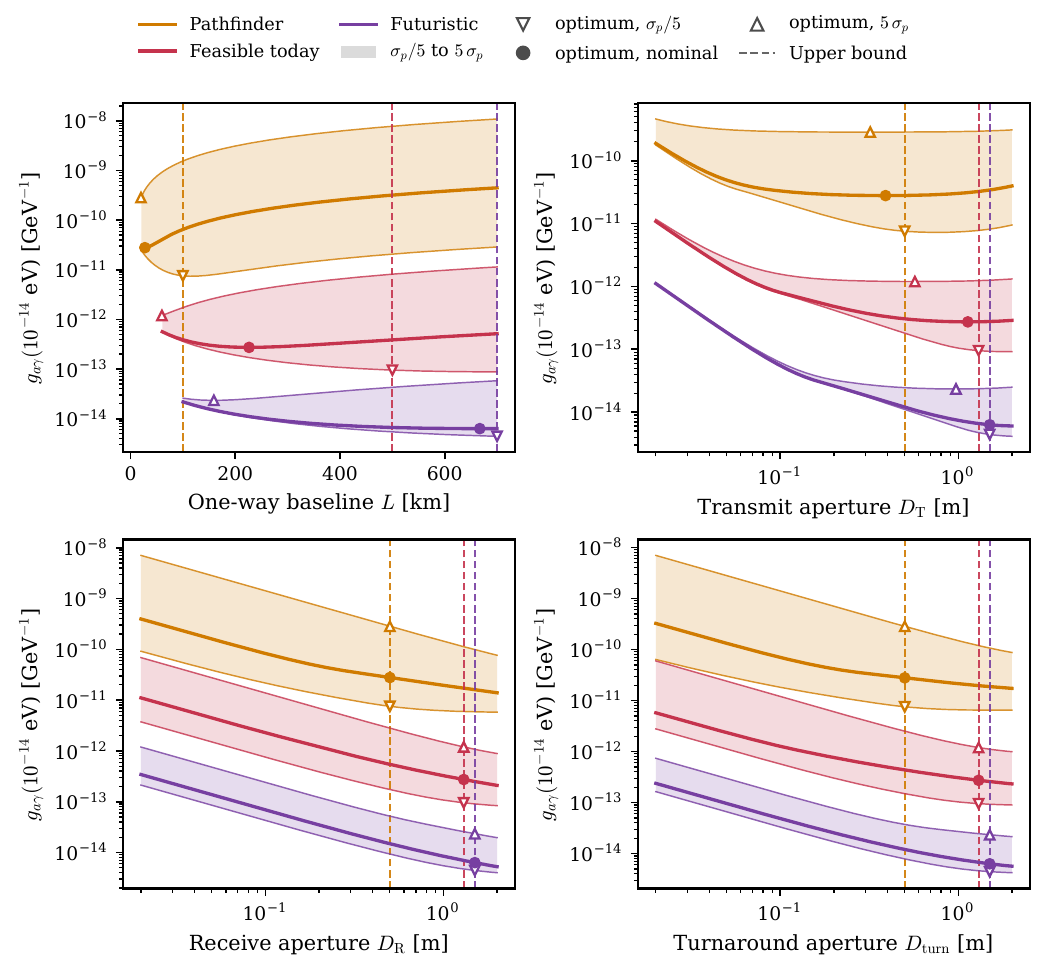}
\caption{Reach at $m_0=10^{-14}$~eV as a function of one geometric parameter. At each
point the other three parameters are re-optimized within the bounds of
Table~\ref{tab:optimization}, including the separation drift of App.~\ref{app:separation}.
Thick lines use the pointing error of Table~\ref{tab:config}, and filled circles mark
the optimized designs of Table~\ref{tab:optimization}. Shaded bands show the reach for a
pointing error between $\sigma_p/5$ (lower edge) and $5\,\sigma_p$ (upper edge), with
$\varepsilon_{\rm turn}=\sigma_p$; open downward and upward triangles mark the
corresponding optimized designs. Dashed vertical lines show the upper bounds of
Table~\ref{tab:optimization}; the scans extend beyond them. Baselines are restricted to
$L\geq2\sigma_{\rm rel}$, where the drift model of App.~\ref{app:separation} applies. At small transmit apertures the beam divergence is set by diffraction rather than by the
pointing, so the bands of the Feasible today and Futuristic configurations close.}
\label{fig:design-scans}
\end{figure*}

Figure~\ref{fig:design-scans} shows how the reach changes when one parameter is varied
while the other three are re-optimized at each point. In the baseline scan, a longer
$L$ increases the axion response but reduces the returned power. The two effects now
balance inside the bounds for all three levels, at $27.5$, $226$ and $667$~km. The
transmit aperture has an optimum inside the bounds for the Pathfinder and Feasible
today levels, at $0.39$ and $1.13$~m. This is slightly above the diffraction optimum
$D_{\rm T}\simeq3.03\sqrt{\lambda L/\pi}$ of Eq.~\eqref{eq:optimal-waist}, because the returning beam
travels $2L$; for the Futuristic level this optimum lies above the $1.5$~m bound. The
receive and turnaround apertures always sit at their upper bound: larger apertures
would still improve the reach, since only $7$--$18\%$ of the returning beam is
collected.

The shaded bands show the effect of the pointing precision, varied from $\sigma_p/5$ to
$5\,\sigma_p$ (with $\varepsilon_{\rm turn}=\sigma_p$), with the geometry re-optimized for
each value. The effect is asymmetric. A pointing error five times larger degrades the
reach at $m_0$ by factors of $10$, $4.4$ and $3.7$ for the Pathfinder, Feasible today
and Futuristic levels. The beam broadening $\propto L\sigma_p$ then dominates, and the
optimum moves to shorter baselines: $20$, $60$ and $159$~km, the first two at the lower
limit $L=2\sigma_{\rm rel}$ of the drift model. If the geometry is instead kept at its
nominal optimum, the loss grows to factors of $15$, $14$ and $9$. A pointing error five
times smaller improves the reach by factors of only $3.7$, $2.9$ and $1.4$, because the
optimum then reaches the bounds on $L$ and on the apertures. The pointing is therefore
the most critical requirement for the Pathfinder: at $5\,\sigma_p=40~\mu$rad its reach,
$2.8\times10^{-10}$~GeV$^{-1}$, would lie above the CAST bound. Feasible today would move
from about a factor of two below to a factor of two above the astrophysical bounds,
while the Futuristic level stays about twenty times below them even at $5\,\sigma_p$. At
small transmit apertures the beam divergence is set by diffraction rather than by the
pointing, and the bands of the Feasible today and Futuristic levels close.

\providecommand{\doi}[1]{\href{https://doi.org/#1}{\nolinkurl{#1}}}
\clearpage
\bibliography{refs}
\end{document}